# Round-Trip Voronoi Diagrams and Doubling Density in Geographic Networks

Matthew T. Dickerson
Middlebury College
Middlebury, VT, USA
Email: dickerso@middlebury.edu

Michael T. Goodrich
Univ. of California, Irvine
Irvine, CA, USA
Email: goodrich@ics.uci.edu

Thomas D. Dickerson
St. Michael's College
Colchester, VT, USA
Email: tdickerson@smcvt.edu

**Abstract**

The *round-trip* distance function on a geographic network (such as a road network, flight network, or utility distribution grid) defines the "distance" from a single vertex to a pair of vertices as the minimum length tour visiting all three vertices and ending at the starting vertex. Given a geographic network and a subset of its vertices called *sites* (for example a road network with a list of grocery stores), a *two-site round-trip Voronoi diagram* labels each vertex in the network with the pair of sites that minimizes the round-trip distance from that vertex. Alternatively, given a geographic network and *two* sets of sites of different types (for example grocery stores and coffee shops), a *two-color round-trip Voronoi diagram* labels each vertex with the pair of sites of different types minimizing the round-trip distance. In this paper, we prove several new properties of two-site and two-color round-trip Voronoi diagrams in a geographic network, including a relationship between the *doubling density* of sites and an upper bound on the number of non-empty Voronoi regions. We show how those lemmas can be used in new algorithms asymptotically more efficient than previous known algorithms when the networks have reasonable distribution properties related to doubling density, and we provide experimental data suggesting that road networks with standard point-of-interest sites have these properties.

## 1 Introduction

A *geographic network* is a graph $G = (V, E)$ that represents a transportation or flow network, where commodities or people are constrained to travel along the edges of that graph. Examples include road, flight, and railroad networks, utility distribution grids, and sewer lines. We assume that the edges of a geographic network are assigned weights, which represent the cost, distance, or penalty of moving along that edge, or some combination of these and other factors, such as scenic or ecological value. The only requirement we make with respect to these weights is that they be non-negative. In this paper, we also restrict our attention to undirected geographic networks.

Since all our edge weights are non-negative, and the edges are undirected, a shortest path exists between each pair of vertices in $G$. The distance, $d(v, w)$ for $v, w \in G$, is defined as the *length* of a *shortest* (i.e., minimum weight) path between $v$ and $w$. This distance function, $d$, is well-defined, and $d(v, w) = d(w, v)$. The *triangle inequality* holds for this path distance $d$.

This observation allows us to define the Voronoi diagram of a geographic network. Formally, we define a geographic network, $G = (V, E)$, to be a set $V$ of *vertices*, a set $E$ of *edges* (which are unordered pairs of distinct vertices), and a *weight function* mapping $E$ to non-negative real numbers. In a road network, this weight function could represent either distance along a road (that is, the Euclidean length of an edge) or the travel time. In the Voronoi diagram problem, we are also given a subset $K \subset V$ of special vertices called *sites*. These are the "post offices" in Knuth's post office problem, but of course they could also be any points of interest (or POIs) such as a schools, hospitals, fire stations, or grocery stores. Each site $v \in K$ is uniquely labeled with a natural number $n(v)$ from 0 to $|K| - 1$, so that we can refer to sites by number. The numbering is also used to resolve ties so that the ordering of sites by distance can be uniquely defined.

The standard first-order graph-theoretic *Voronoi diagram* [14] of $G$ is a labeling of each vertex $w$ in $V$ with the number, $n(v)$, of the vertex $v$ in $K$ that is closest to $w$. All the vertices with the same label, $n(v)$, are said to be in the *Voronoi region* for $v$. Intuitively, if a site $v$ in $K$ is considered a post office, then the Voronoi region for $v$ consists of all the homes that *ought* to be in $v$'s zip code. (See Fig. 1.)

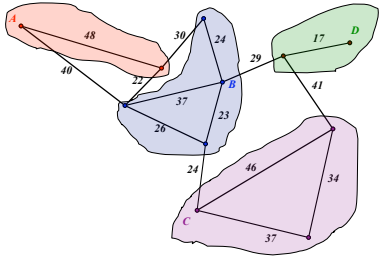

Figure 1: An example graph-theoretic Voronoi diagram with sites A,B,C,D.

Mehlhorn [14] shows that the graph-theoretic Voronoi diagram of a graph $G$, having $n$ vertices and $m$ edges, can be constructed in $O(n \log n + m)$ time. A similar algorithm is given by Erwig [10]. At a high level, these algorithms perform $n$ simultaneous runs of Dijkstra's single-source shortest-path algorithm [9] (see also [6, 11]). In this paper, however, we are not interested in these types of *single-site* Voronoi diagrams.

## 1.1 Round-Trip Distance

In a number of applications, we may be interested in labeling the vertices of a geographic network, $G$, with more information than just their single nearest neighbor from the set of sites, $K$. We may wish, for instance, to label each vertex $v$ in $G$ with the names of the $C$ closest sites in $K$, for some $C \leq |K|$. For example, the sites in $K$ may be fire stations, and we may wish to know the three closest fire stations for each house in our network, just in case there is a three-alarm fire at that location.

For many applications, "closest" among a set of neighbors should instead be defined by the *round-trip* or *tour* distance. (For $C = 2$, and for point sites on the Euclidean plane, this distance was referred to in [4] as the "perimeter" distance, in reference to the perimeter of a triangle, since shortest paths are straight edges. In this paper, although we also focus on the case $c = 2$, we will refer to this function as the "round-trip" distance in reference both to road networks and to the more general case of $C \geq 2$.) In this notion of distance, we want to take a single trip, starting and ending at our "home" location and visiting two (or more) distinguished sites. Such distances correspond to the work that would need to be done, for example, by someone who needs to leave their house, visit multiple sites to run a number of errands, and then return home. Some motivating examples include the following:

- Some legal documents require the signatures of multiple witnesses and/or notaries in order to be executed, so we may need to travel to multiple locations to get them all.

- Some grocery stores place a limit on the amount of special "loss leader" sale items one can purchase in a single visit, so we may need to visit multiple stores to get enough of such items needed for a big party.

- A celebrity just out of rehab may wish to get multiple community service credits in a single trip, for instance, by tutoring students at an educational institution and speaking at an alcohoics anonymous meeting at a religious institution, all on the same day.

In each case, we are likely to want to optimize our travel time to visit all the sites of interest as quickly as possible. Thus, we focus in this paper on the construction of *multi-site* Voronoi diagrams for round-trip distance functions. (See Fig. 2.)

Alternatively, we may have a number of different kinds of sites, such as gas stations, grocery stores, and coffee houses, and we are interested in the three that are closest to each house, in terms of how one could visit all three types of sites in a single trip from home. Thus we are also interested in *multi-color* Voronoi diagrams, where each type of site (such as coffee houses and grocery stores) is represented with a different color.

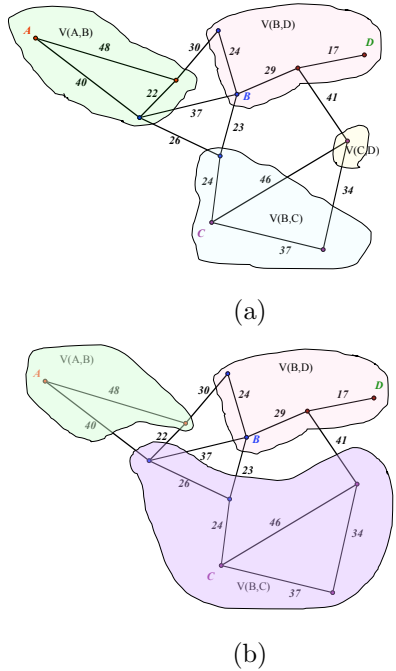

Figure 2: (a) An example graph-theoretic two-site sum function Voronoi diagram of the same graph from Fig. 1. (b) An example graph-theoretic two-site round-trip function Voronoi diagram of the same graph as in Fig. 1.

## 1.2 Related Prior Work

Unlike prior work on graph-theoretic Voronoi diagrams, there is a abundance of prior work for geometric Voronoi diagrams. It is beyond the scope of this paper to review all this work and its applications. We refer the interested reader to any excellent survey on the subject (e.g., see [1]) and we focus here on previous work on multi-site geometric Voronoi diagrams and on graph-theoretic Voronoi diagrams.

Lee [13] studies $k$-nearest neighbor Voronoi diagrams in the plane, which are also known as "order-$C$ Voronoi diagrams." These structures define each region, for a site $p$, to be labeled with the $C$ nearest sites to $p$. These structures can be constructed for a set of $n$ points in the plane in $O(n^2 + C(n-C)\log^2 n)$ time [5]. Due to their computational complexity, however, order-$C$ Voronoi diagrams have not been accepted as practical solutions to $C$-nearest neighbor queries. Patroumpas *et al.* [15] study methods for performing $C$-nearest neighbor queries using an approximate order-$C$ network Voronoi diagram of points in the plane, which has better performance than its exact counterpart.

Two-site distance functions and their corresponding Voronoi diagrams were introduced by Barequet, Dickerson, and Drysdale [4]. A two-site distance function is measured from a point to a pair of points. In Euclidean space, it is a function $D_f : \mathbf{R}^2 \times (\mathbf{R}^2 \times \mathbf{R}^2) \to \mathbf{R}$ mapping a point $p$ and a pair of points $(v, w)$ to a non-negative real number. The sum function, $D_S$, results in the same Voronoi diagram as the 2-nearest neighbor (order-2) Voronoi diagram, but the authors considered a number of other combination rules as well including *area* and *product*. The complexity of the *round-trip* two-site distance function Voronoi diagram was left by [4] as an open problem, and remains open.

As we mentioned above, single-site graph-theoretic Voronoi diagrams were considered by Mehlhorn [14], who presented an algorithm running in $O(n \log n + m)$ time. Bae and Chwa [2, 3] study hybrid schemes where distance is defined by a graph embedded in the plane and distance is defined by edge lengths.

As far as multi-site queries are concerned, Safar [16] studies k-nearest neighbor searching in road networks, but he does so using the first-order Voronoi diagram, rather than considering a multi-site Voronoi diagram for geographic networks. Likewise, Kolahdouzan and Shahabi [12] also take the approach of constructing a first-order Voronoi diagram and searching it to perform $C$-nearest neighbor queries. Instead, de Almeida and Güting [7] compute $C$-nearest neighbors on the fly using Dijkstra's algorithm. None of these methods actually construct a multi-site or multi-color graph-theoretic Voronoi diagram, however, and, to the the best of our knowledge, there is no previous paper that explicitly studies multi-site or multi-color Voronoi diagrams on graphs. In [8], Dickerson and Goodrich study two-site Voronoi diagrams in graphs, but without employing any techniques that could improve running times beyond repeated Dijkstra-like algorithms.

## 1.3 Our Results

In this paper, we focus on two-site and two-color Voronoi diagrams on graphs using the round-trip function $D_P$ for defining these concepts, although we also discuss the sum distance function, $D_S$, as well. In particular, for a vertex $p$ or a point $p$ on an edge $e$ and a pair of sites $v, w$, our two-site distance functions are defined as follows:

$$\begin{aligned} D_S(p, (v, w)) &= d(p, v) + d(p, w) \\ D_P(p, (v, w)) &= d(p, v) + d(p, w) + d(v, w) \end{aligned}$$

The sum function can easily be extended from 2 to $k$ sites: $D_S(p, (v_1, \ldots, v_k)) = \sum_{1 \leq i \leq k} d(p, v_i)$. Note that with $k$-site distance functions, we also have a similar rule for breaking ties in distances.

We prove several new properties of two-site round-trip distance function Voronoi diagrams on geographic networks, and make use of these properties to provide a new family of algorithms for computing these diagrams. We extend our proofs for the two-color variant, which is arguably more applicable than the one-color variant. (Though as noted above, there are cases when one might wish to visit several grocery stores on one trip, it is easier to imagine a case where we want the shortest tour visiting both a grocery store and a post office.)

One property we explore relates to the *doubling densities* of various types of POI sites on a geographic networks. The doubling density of a class of sites from a vertex $v$ is the number of sites of that type within twice the distance from $v$ as the closest site to $v$ of that type. The run-times of our algorithms depend in part on the average doubling density of various sites from other sites. They also depend on the related density of the total number of edges within twice the distance from one site to the nearest other site of that type. (This latter property could be thought of as a different kind of doubling density.) We will prove a property that allows us to prune our search based on doubling distances, and will also provide experimental results about the doubling densities of various POIs on a set of states.

The algorithms have run times whose expected case is asymptotically faster than the algorithm of [8] under realistic assumptions of how sites are distributed in the network. Finally, we show how to extend two-site Voronoi diagrams to multi-site and multi-color diagrams, under the sum function, while only increasing the running time by a factor of $C$, where $C$ is the multiplicity we are interested in.

## 2 Constructing Graph-Theoretic Voronoi Diagrams

In this section, we review the approach of Mehlhorn [14] and Erwig [10] for constructing a (single-site) graph-theoretic Voronoi diagram of a graph $G$, having $n$ vertices and $m$ edges, which runs in $O(n \log n + m)$ time, and, for completeness, we also review the two-site sum function algorithm of [8], but with one minor correction.

Given a geographic network, $G = (V, E)$, together with a set of sites, $K \subseteq V$, and a non-negative distance function on the edges in $E$, the main idea for constructing a graph-theoretic Voronoi diagram for $G$ is to conceptually create a new vertex, $a$, called the *apex*, which was originally not in $V$, and connect $a$ to every site in $K$ by a zero-weight edge. We then perform a single-source, shortest-path (SSSP) algorithm from $a$ to every vertex in $G$, using an efficient implementation of Dijkstra's algorithm. Intuitively, this algorithm grows the Voronoi region for each site out from its center, with the growth for all the sites occurring in parallel. Moreover, since all the Voronoi regions grow simultaneously and each region is contiguous and connected by a subgraph of the shortest-path tree from $a$, we can label vertices with the name of their Voronoi region as we go.

In more detail, the algorithm begins by labeling each vertex $v$ in $K$ with correct distance $D[v] = 0$ and every other vertex $v$ in $V$ with tentative distance $D[v] = +\infty$, and we add all these vertices to a priority queue, $Q$, using their $D$ labels as their keys. In addition, for each vertex $v$ in $K$, we label $v$ with the name of its Voronoi region, $R(v)$, which in each case is clearly $R(v) = n(v)$. In each iteration, the algorithm removes a vertex $v$ from $Q$ with minimum $D$ value, confirming its $D$ label and $R$ label as being correct. It then performs a *relaxation* for each edge $(v, u)$, incident to $v$, by testing if $D[v] + w(v, u) < D[u]$. If this condition is true, then we set $D[u] = D[v] + w(v, u)$, updating this key for $u$ in $Q$, and we set $R(u) = R(v)$, to indicate (tentatively) that, based on what we know so far, $u$ and $v$ should belong to the same Voronoi region. When the algorithm completes, each vertex will have its Voronoi region name confirmed, as well as the distance to the site for this region. Since each vertex is removed exactly once from $Q$ and each key is decreased at most $O(m)$ times, the running time of this algorithm is $O(n \log n + m)$ if $Q$ is implemented as a Fibonacci heap. In addition, note that this algorithm "grows" out the Voronoi regions in increasing order by distance from the apex, $a$, and it automatically stops the growing of each Voronoi region as soon as it touches another region, since the vertices in an already completed region are (by induction) closer to the apex than the region we are growing.

## 2.1 Two-site Distance Functions on Graphs

As mentioned above, the two-site sum function Voronoi diagram is equivalent to the second order two-nearest neighbor Voronoi diagram, which identifies for each vertex $v$ in our graph, $G$, the two nearest sites to $v$. It follows that the two-site Voronoi diagram is equivalent to the two-nearest neighbor Voronoi diagram for a set of points in the plane or a graph.

Intuitively, the algorithm of [8] for constructing a two-site Voronoi diagram under the sum function is to perform a Dijkstra single-source shortest-path (SSSP) algorithm from each site, in parallel, but visit each vertex twice—once for each of the two closest sites to that vertex.

More specifically, we begin by labeling each vertex $v$ in $K$ with correct first-neighbor distance $D_1[v] = 0$ and every other vertex $v$ in $V$ with tentative first-neighbor distance $D_1[v] = +\infty$, and we add all these vertices to a priority queue, $Q$, using their $D_1$ labels as their keys. We also assign each vertex $v \in V$ (including each site in $K$) its tentative second-neighbor distance, $D_2[v] = +\infty$, but we don't yet use these values as keys for vertices in $Q$. In addition, for each vertex $v$ in $K$, we label $v$ with the name of its first-order Voronoi region, $R_1(v)$, which in each case is clearly $R_1(v) = n(v)$. In each iteration, the algorithm removes a vertex $v$ from $Q$ with minimum key. How we then do relaxations depends on whether this key is a $D_1$ or $D_2$ value.

- Case 1: The key for $v$ is a $D_1$ value. In this case we confirm the $D_1$ and $R_1$ values for $v$, and we add $v$ back into $Q$, but this time we use $D_2[v]$ as $v$'s key. We then perform a *relaxation* for each edge $(v, u)$, incident to $v$, according to the following test:

  **Relaxation**$(v, u)$:
      **if** $u$ has had its $R_2$ label confirmed **then**
        Return (for we are done with $u$).
      **else if** $u$ has had its $R_1$ label confirmed **then**

        **if** $R_1(v) \neq R_1(u)$ and $D_1[v] + w(v,u) < D_2[u]$ **then**
           Set $D_2[u] = D_1[v] + w(v,u)$
           Set $R_2(u) = R_1(v)$
        **if** $D_2[v] + w(v,u) < D_2[u]$ **then**
           Set $D_2[u] = D_2[v] + w(v,u)$
           Set $R_2(u) = R_2(v)$.
    **else**
        **if** $D_1[v] + w(v,u) < D_1[u]$ **then**
           Set $D_1[u] = D_1[v] + w(v,u)$
           Set $R_1(u) = R_1(v)$.

In addition, if the $D_1$ or $D_2$ label for $u$ changes, then we update this key for $u$ in $Q$. Moreover, since we are confirming the $D_1$ and $R_1$ labels for $v$, in this case, we also do a reverse relaxation for each edge incident to $v$ by calling **Relaxation**$(u,v)$ on each one.

- Case 2: The key for $v$ is a $D_2$ value. In this case we confirm the $D_2$ and $R_2$ values for $v$, and we do a relaxation for each edge $(v,u)$, incident to $v$, as above (but with no reverse relaxations).

When the algorithm completes, each vertex will have its two-site Voronoi region names confirmed, as well as the distance to each of its two-nearest sites for this region. For the analysis of this algorithm, first note that no vertex will be visited more than twice, since each vertex is added to the queue, $Q$, twice—once for its first-order nearest neighbor and once for its second-order nearest neighbor. Moreover, once a vertex is added to $Q$, its key value is only decreased until it is removed from $Q$. Thus, this algorithm requires $O(n \log n + m)$ time in the worst cast when $Q$ is implemented using a Fibonacci heap, where $n$ is the number of vertices in $G$ and $m$ is the number of edges. By the same reasoning, the priority queue $Q$ won't grow larger than $O(n)$ during the algorithm, so the space required is $O(n)$.

# 3 Properties of Round-Trip Voronoi Diagrams on Graphs

Using the sum distance function for a two-site graph-theoretic Voronoi diagram allows us to label each vertex in $G$ with its two nearest neighbors. Such a labeling is appropriate, for example, for fire stations or police stations, where we might want agents from both locations to travel to our home, or if we need to take separate trips to different sites. If instead we want to leave our home, travel to two nearby sites on the same trip, and return home, then we will need to use the round-trip function. Before presenting a new algorithm for this function, we first prove several properties of the round-trip distance function diagram on graphs.

Our first lemma is relatively straightforward, but provides an important property for pruning searches in our algorithms.

**Lemma 1** *Let $v$ be any vertex in a geographic network $G$, $(s,t)$ a pair of sites in $G$ minimizing the round-trip distance function $D_P$ from $v$. Then for any sites $p, q$ in $G$:*

$$d(v,s) \leq (d(v,p) + d(v,q) + d(p,q))/2$$
$$d(v,t) \leq (d(v,p) + d(v,q) + d(p,q))/2$$

**Proof of Lemma 1:** By assumption, $D_P(v,(s,t)) \leq D_P(v,(p,q))$. By the triangle inequality, $2d(v,s) \leq d(v,s) + d(v,t) + d(s,t)) = D_P(v,(s,t))$. Combining these, we get, $d(v,s) \leq \frac{1}{2} D_P(v,(p,q)) = \frac{1}{2}(d(v,p) + d(v,q) + d(p,q))$. The argument for $d(v,t)$ is symmetric. **End Proof.**

What this means is that if we know of some tour from vertex $v$ through sites $p$ and $q$—that is, we have a *candidate* pair $(p,q)$ to minimize the round-trip distance from $v$— then our algorithms can safely ignore any other site $s$ that is further from $v$ than $\frac{1}{2} D_P(v,(p,q))$ because $s$ cannot be a part of a pair that minimizes the round-trip distance from $v$.

This lemma combined with the triangle inequality $d(p,q) \leq d(v,p) + d(v,q)$ leads to the following corollary, which is a weaker condition, but one easily implementable as a pruning technique on a SSSP search.

**Corollary 2** *Let $p, q$ be the two sites closest to some vertex $v$ under normal graph distance, and $(s,t)$ the pair of sites minimizing the round-trip function $D_P$ from $v$. Then $d(v,s) \leq d(v,p) + d(v,q)$ and $d(v,t) \leq d(v,p) + d(v,q)$.*

The following *double distance* lemma provides a similar but less obvious condition that can also be used for pruning.

**Lemma 3** *(Doubling Distance Property) For any pair of sites $s, t$ in a geographic network $G$, if there exists any other sites $p, q \in G$ such that $d(s,t) > 2d(s,p)$ and $d(s,t) > 2d(t,q))$, then $(s,t)$ cannot minimize the round-trip distance function for* any *vertex $v \in G$.*

**Proof of Lemma 3 (by contradiction):** (See Figure 3.) Assume that there is some vertex $v$ such that $(s,t)$ is the closest pair of sites in the round-trip distance—that is, $v$ is in the Voronoi region for $(s,t)$. Assume also that there are sites $p, q \in G$ such that $d(s,t) > 2d(s,p)$ and $d(s,t) > 2d(t,q))$. Without loss of generality, let $d(v,s) \leq d(v,t)$. (Otherwise reverse the role of $s$ and $t$.) We now consider the round-trip distance $D_P(v,(s,p))$. Applying the triangle inequality, we get: $D_P(v,(s,p)) = d(v,s) + d(s,p) + d(p,v) \leq d(v,s) + d(s,p) + (d(v,s) + d(s,p))$. By assumption, $2d(s,p) < d(s,t)$ and $d(v,s) \leq d(v,t)$ and thus: $D_P(v,(s,p)) < d(v,s) + d(v,t) + d(s,t) = D_P(v,(s,t))$, contradiction our assumption that $(s,t)$ is the closest pair to $v$. **End Proof**

Note that Lemma 3 holds even if if $s$ and $t$ both meet the condition of Lemma 1—that is, even if for some vertex $v$, $s$ and $t$ are both closer to $v$ than $\frac{1}{2} D_P(v,(p,q))$ for all sites $p, q$, the pair $(s,t)$ cannot minimize the round-trip distance from $v$. Thus if the conditions of Lemma 3 hold, it follows immediately that the Voronoi region for $(s,t)$ is empty in the two-site round-trip distance function Voronoi diagram.

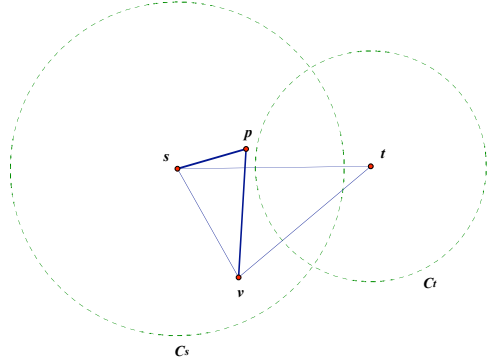

Figure 3: Illustrating the proof of Lemma 3. The edges represent shortest paths, not single edges.

We now state a final property of round-trip function Voronoi diagrams on graphs.

**Lemma 4** *Let $s$ be any site in a geographic network $G$, and $v, w$ any vertices in $G$ such that a shortest path from $s$ to $v$ goes through $w$. If there exist any sites $p, q \in G$ such that $d(w, s) > \frac{1}{2}(d(w, p) + d(w, q) + d(p, q))$ then $s$ cannot be part of a nearest round-trip pair to $v$.*

**Proof of Lemma 4:** ( See Figure 4.) For any $t$ we know the following by the triangle inequality and the fact that $w$ is on the shortest path from $s$ to $v$:

$$d(v, t) + d(t, s) \geq d(v, w) + d(w, s) \tag{1}$$

Also by assumption

$$2d(w, s) > d(w, p) + d(w, q) + d(p, q). \tag{2}$$

Using Equations 1 and 2, we now show that $D_P(v, (p, q)) < D_P(v, (s, t))$ for $s$ and any other site $t$.

$$\begin{aligned}
D_P(v, (p, q)) &= d(v, p) + d(v, q) + d(p, q) \\
D_P(v, (p, q)) &\leq [d(v, w) + d(w, p)] + [d(v, w) + d(w, q)] + d(p, q) \\
D_P(v, (p, q)) &\leq 2d(v, w) + d(w, p) + d(w, q) + d(p, q) \\
D_P(v, (p, q)) &< 2d(v, w) + 2d(w, s) \\
D_P(v, (p, q)) &< 2d(v, t) + 2d(t, s) \\
D_P(v, (p, q)) &< d(v, t) + d(t, s) + d(v, s) \\
D_P(v, (p, q)) &< D_P(v, (s, t))
\end{aligned}$$

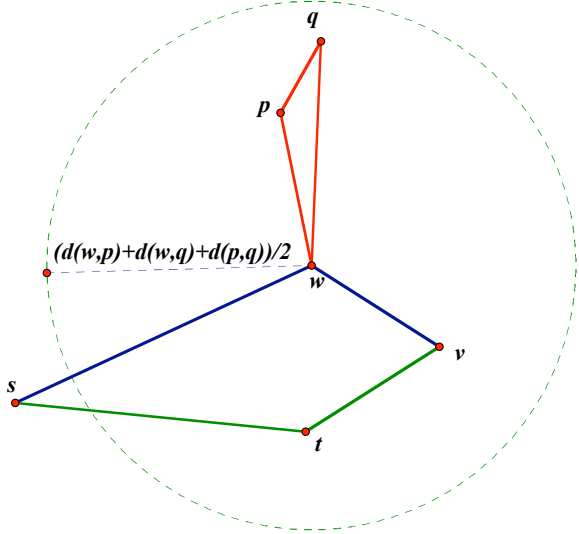

Figure 4: Illustrating the proof of Lemma 4. The edges represent shortest paths in the graph, and not single edges. The edges $(s, w)$ and $(w, v)$ represent (by assumption) the *shortest* path from $s$ to $v$.

**End Proof.**

We could rephrase this in the contrapositive, in a form similar to that of Lemma 1. Let $v$ be any vertex in a geographic network $G$, $(s, t)$ a pair of sites in $G$ minimizing the round-trip distance function $D_P$ from $v$, $w$ a vertex on a shortest path from $s$ to $v$, and $p, q$ any sites in $G$, then $d(w, s) \leq (d(w, p) + d(w, q) + d(p, q))/2$. What this Lemma means is that even if the pair of sites $(s, t)$ meet the condition of Lemma 1 for some vertex $v$—that is, $s$ is a candidate site to be part of the closest pair to $v$—if the shortest path from $s$ to $v$ goes through some vertex $w$ for which $s$ does not meet that condition, then not only is $s$ not part of a closest pair for $w$, $s$ also cannot be part of a closest round-trip pair to $v$. Together, these three lemmas are sufficient to prove the correctness of the algorithms in the following section.

## 3.1 Two-Color Variants

These lemmas can all be extended to apply to the two-color variant. The two color versions are given below. In the follow lemmas, we let $G = (V, E, S, T)$ be a geographic network, with $S \subset V$ and $T \subset V$ two disjoint sets of sites (of different colors). The two-color round-trip distance is from a vertex in $V$ to a pair of sites $(s, t)$ with $s \in S$ and $t \in T$. The proofs of these lemmas are directly analogous to the proofs above.

**Lemma 5** *Let $v$ be any vertex in $G$. Let $s \in S$ and $t \in T$ be a pair of sites such that $(s, t)$ minimizes the two-color round-trip distance function $D_P$ from $v$. Let*

$p \in S$ and $q \in T$ be sites in $G$. Then:

$$d(v,s) \leq (d(v,p) + d(v,q) + d(p,q))/2$$
$$d(v,t) \leq (d(v,p) + d(v,q) + d(p,q))/2$$

**Corollary 6** *Let $p \in S$ and $q \in T$ be the sites in $S$ and $T$ respectively that are closest to some vertex $v$ under normal graph distance, and let $(s,t)$ (with $s \in S$ and $t \in T$) be the pair of sites minimizing the two-color round-trip function $D_P$ from $v$. Then $d(v,s) \leq d(v,p) + d(v,q)$ and $d(v,t) \leq d(v,p) + d(v,q)$.*

**Lemma 7** *For any pair of sites $s \in S$ and $t \in T$ in a geographic network $G$, if there exists any other sites $p \in T$ and $q \in S$ such that $d(s,t) > 2d(s,p)$ and $d(s,t) > 2d(t,q)$), then $(s,t)$ cannot minimize the round-trip distance function for any vertex $v \in G$.*

**Lemma 8** *Let $s$ be any site in a geographic network $G$, and $v,w$ any vertices in $G$ such that a shortest path from $s$ to $v$ goes through $w$. If there exist any sites $p \in S$ and $q \in T$ such that $d(w,s) > \frac{1}{2}(d(w,p) + d(w,q) + d(p,q))$ then $s$ cannot be part of a nearest round-trip pair to $v$.*

## 4 Round-Trip Voronoi Diagram Algorithms

We now provide algorithms to compute the round-trip function Voronoi diagram for a geographic network and set of sites $G = (V, K, E)$. Specifically, the algorithm labels each vertex $v \in V$ with a pair of sites in $K$ minimizing the two-site round-trip distance function from $v$.

### 4.1 A Brute Force Algorithm

An algorithm for this problem was first presented in [8]. The algorithm, in Step 1, performs a complete SSSP algorithm on $G$ from each of the $k$ sites in $K$. (Unlike in the Sum function algorithm above, these searches do not need to be interleaved—that is, performed in parallel—as the algorithm searches the entire graph from each site.) It records the distances from $v$ to every site in $K$, and then creates a table of distances between all pairs of sites $(p,q) \in K$, allowing constant time access to $d(p,q)$. Then, in Step 2, for each vertex $v \in V$ and each pair of sites $(p,q) \in S$, the algorithm explicitly computes the round-trip distance

$$D_P(v,(p,q)) = d(v,p) + d(v,q) + d(p,q)$$

and labels each $v$ with pair $(p,q)$ minimizing this function.

This brute force approach uses the SSSP algorithm to efficiently compute all distances between pairs of vertices, and then explicitly compares all round-trip distances. The algorithm requires $O(k^2 n + km + kn \log n)$ time and $O(nk)$ space when we implement it using Fibonacci heaps as discussed above.

## 4.2 Improving the Brute Force Method: a Revised Algorithm

We now show how the properties of the previous section can be used to prune the search depth of the brute force algorithm of [8]. Our new algorithm has three steps, or phases.

Step 1 corresponds to Step 1 of the brute force approach above, except that we interleave the SSSP searches (as is done with the sum function) and we bound the number of sites that visit each vertex using some value $B$. In practice, this bounds the SSSP search outward from each site in $K$. The specific value of $B$–possibly determined as a function of $n, m, k$ –will be described in the next section; the algorithm is correct regardless of the value of $B$, but its run time will depend on $B$. Ideally, the SSSP of Step 1 provides enough information for most (or all) of the vertices to determine the pair of sites minimizing the round-trip distance. However the pruning may result in some vertices having incomplete information.

In Step 2, we need to complete information for each of these vertices that still have incomplete information by preforming an addition SSSP search outward from that vertex until it reaches all the sites satisfying Lemma 1. In particular, the smaller the value of $B$, the less work is done in Step 1, but the more potential work will need to be done in Step 2.

By Step 3, we have all the distance information needed to compute $D_P(v, (s, t))$ for all pairs of sites $(s, t)$ satisfying Corollary 2, and Lemmas 3 and 4. We need only explicitly compute these distances from the information computed in Steps 1 and 2. Note that the pruning of Step 1 reduces not only the time required by each SSSP search, but also the number of explicit distances computed.

We start with the basic three-phase revised algorithm to compute the round-trip distance function two-site Voronoi diagram on a graph $G = (V, K, E)$.

- **Step 1:** Perform parallel Dijkstra SSSP algorithm from each site $p \in K$. For each vertex $v \in V$, record the distances from the first $B+1$ sites whose SSSP search visits $v$. Any subsequent search (after the $B+1^{st}$) visiting $v$ is not recorded and the search is terminated. (As we will show, the result of Step 1 is that for each vertex, we have a list of the $B + 1$ closest sites in sorted order.

- **Step 2:** For each vertex $v \in V$, let $p, q$ be two closest sites in $K$, and compute $d_v = d(v, p) + d(v, q)$. By Corollary 2, no site further than $d_v$ from $v$ is a candidate to be part of a pair minimizing the round-trip distance from $v$. So we consider two cases:

  **case i:** If the final site $p$ on the sorted list of $B + 1$ closest sites to $v$ is further from $v$ than $d_v$, then we have found all sites closer to $v$ than $d_v$ and no work needs to be done on vertex $v$ in this Step; the list of sites at $v$ contains all possible candidate sites that could be part of a closest pair in the round-trip function.

**case ii:** If the final site on the sorted list for $v$ is not further than $d_v$, then we cannot guarantee that $v$ was visited by all the candidate sites. In this case, we perform a SSSP algorithm from $v$ and halt when we reach any vertex further from $v$ than $d_v$. (Note that this is done also for those vertices that are also sites in $K$. Since $d_v$ is at least as great as twice the distance from $v$ to its nearest site, we will compute distances between all pairs of sites satisfying Lemma 3.)

- **Step 3:** For each vertex $v \in V$, compute $D_P(v,(s,t)) = d(v,s) + d(v,t) + d(s,t)$ for all sites $s,t$ for which $d(v,s)$ and $d(v,t)$ are stored at $v$ and $d(s,t)$ is stored at either $s$ or $t$. (If $d(s,t)$ was not computed, then $(s,t)$ is not a candidate pair and may be ignored.) Store at $v$ the pair $(s,t)$ minimizing $D_P(v,(s,t))$.

### 4.2.1 Correctness

Since the first $B+1$ SSSP searches that reach any vertex will continue through the vertex, by induction each vertex is guaranteed to be reached by the SSSP from at least its closest $B+1$ sites in Step 1. In Step 2, therefore, by looking at the first two and the last site in the list for each vertex $v$, we can determine if all sites meeting Corollary 2 have visited $v$. If not, then an SSSP from $v$ (in case ii) will reach those sites. So by Corollary 2 and Lemma 4, any sites $s,t$ for which the algorithm does not explicitly computer $D_P(v,(s,t))$ cannot be a candidate to minimize the round-trip distance from $v$.

### 4.2.2 Worst Case Analysis

We now analyze the algorithm. In Step 1, we visit each vertex $B+1$ times. (If a search arrives at a vertex $v$ that has already been visited $B+1$ times, we count that work to the edge along which the SSSP came to $v$.) An edge can be traversed at most $B+1$ times from the vertices on each end, for a total of $O(B)$ visits. So Step 1 requires $O(Bm + Bn \log n)$ time because we are overlapping $B$ SSSP searches. We are storing $B+1$ sites and distances at each vertex, as well as a list of $O(k^2)$ distances between each pair of sites, so the space required is $O(Bn + m + k^2)$.

In step 2, we need to store distances between pairs of sites $s,t$ that are candidates to minimize round-trip distance for some vertex. If we use a table, we need worst case $O(k^2)$ space with $O(1)$ time access for any pair $(s,t)$. We also store distances between vertex $v$ and its candidate sites in sorted order; there are at most $B$ sites per list in Step 1, and though in Step 2 the lists can grow to size $O(k)$ we only need to store one list at a time, and so space required is $O(Bn + m + k^2)$.

In step 2, if for a vertex $v$, the $B+1$ vertices on its list includes all the sites within distance $d_v$, then we are in case i, and the total amount of work for that vertex in step 2 is $O(1)$ and in step 3 is $O(B^2)$ to explicitly compute all possible round-trip distances of candidate pairs (since for each pair of sites $(s, t)$ the distance $d(s, t)$ has already been computed and can be retrieved in $O(1)$ time.) The total run time for these sites is thus $O(B^2 n)$.

For the rest of the vertices $v$, those in case ii, we must do a new SSSP from $v$. This requires $O(m + n \log n)$ time per vertex for the search and $O(k^2)$ time per vertex to look at all pairs of sites. Let $A$ be the number of sites processed in case ii. The run time for all of them is $O(Am + An \log n + Ak^2)$.

The overall run time is thus $O((A + B)(m + n \log n) + B^2 n + Ak^2)$ and the space required is $O(nB + m + k^2)$.

In the next section, we formalize this and also provide some experimental data on values of $A$ and $B$. First, however, we provide a further revision showing how for many real world networks such as road networks, we can make fuller use of Lemma 1 for an algorithm whose run time is significantly better.

### 4.3 Further Revisions: a Dynamic Variation

It is possible that we can further reduce the depth of our SSSP searches, and thus the number of candidate pairs examined in our algorithm. Lemma 1 gives a stronger condition than Corollary 2 that must be met by any site that is a candidate to minimize the round-trip distance from a vertex $v$.

Specifically, instead of using a static bound that prunes the depth of our searches in Step 1, and then simply computing the distance from vertex $v$ to its two nearest sites, we would like to keep an updated minimal value of $D_P(v, (s, t))$ for *all* sites $s, t$ whose SSSP searches have visited $v$. By Lemmas 1 and 4, we can then prune any search that reaches $v$ from any site further away than the minimum value of $\frac{1}{2} D_P(v, (s, t))$.

Unfortunately, using this stronger condition requires that we dynamically update the minimum value of $D_P(v, (s, t))$ which in turn requires that we precompute or preprocess the values of $d(s, t)$ for all pairs of sites meeting the condition of Lemma 3. This leads to the following two-step algorithm.

- **Step 1:** Perform a SSSP algorithm from each site $p \in K$, terminating the search at the first vertex whose distance from $p$ is greater than $2d(p, q)$ where $q$ is the closest other site to $p$ (discovered in the SSSP). Store the values of $d(p, q)$ for all pairs of sites reached in all of the searches.

- **Step 2:** Perform interleaved SSSP searches from each site $p \in K$, as in Step 1 of the previous algorithm. At each vertex $v$, store the sites $s$ whose searches reach $v$ along with the distance $d(v, s)$. Using this information and the table from Step 1, once a second site search has visited $v$, also compute and maintain the distance $D_P(v, (s, t)) = d(v, s) + d(v, t) + d(s, t)$ that minimizes this function among all pairs of sites $s, t$ which have visited $v$ (as well as the pair $(s, t)$ minimizing that distance). Terminate the search from any site farther from $v$ than $\frac{1}{2} D_P(v, (s, t))$ for the minimum value of $D_P(v, (s, t))$ seen so far.

### 4.3.1 Worst Case Analysis

In the worst case, Step 1 will require $O(m + n \log n)$ time and $O(n + m)$ space for each SSSP for a total of $O(km + kn \log n)$ time, plus an extra $O(k^2)$ space to store the table of distances between pairs of sites, for a total of $O(k^2 + m + n)$ space.

Similarly, in the worst case in Step 2, each of the $k$ SSSP algorithm may require $O(m + n \log n)$ time, but since the searches are interleaved we may need extra $O(nk)$ space to have $k$ searches active at once. We also need to compute $O(k^2)$ distances at each vertex in the worst case, but $k \leq n$ and so we have a total of $O(km + kn \log n)$ time and $O(nk + m)$ space.

As we will see in the following section, however, road networks and many types of POI sites have properties that result in a much more efficient algorithm.

## 4.4 The Two-Color Variant

The algorithms of the previous section can be extended to the two-color variant, where for each vertex $v$ we want to find the pair of sites (or POIs) of two *different* types–say a grocery store and a post office–that minimize the distance of the shortest round-trip from $v$. The same basic approaches of both the revised algorithm and the dynamic variant of the revised algorithm work for the two-color version. Lemmas 5, 7, and 8 suffice as proof.

Other than the obvious change that the two-color versions of the algorithms compute and minimize the round-trip distances to pairs of sites of different types, there are only two other primary changes that are necessary. In the first stage, we still perform the interleaved SSSP algorithms from all sites (of both types). However at each vertex $v$ we store separate lists of the sites of the two different types that visit $v$. This doubles the worst-case memory requirement.

Second, the application of Lemma 7 two-color variant is slightly different than that of Lemma 3 to the standard round-trip distance function. In the dynamic version we need to pre-compute only the distances from sites of one type to sites of the other. In particular, if our two sets of sites are $S$ and $T$, we need to compute the distance from each $t \in T$ to all the sites in $S$ no more than twice the distance of the closest site in $S$ to $t$, and symmetrically from each $s \in S$ to all the sites in $T$ no more than twice the distance of the closest site in $T$ to $s$.

In terms of run-times, what this means for the two-color variant of the round-trip distance function Voronoi diagram is that we care about the *doubling density* of sites in $T$ with respect to sites in $S$ and vice versa–rather than the doubling density of sites in one set to other sites in the same set, as is the case with the standard round-trip two-site distance function.

# 5 Empirical Analysis on Doubling Density and Dynamic Pruning on Road Networks

For the two-color version of the problem we ran experiments on road networks from 14 different U.S. states: CT, HI, IL, IN, LA, MA, MD, ME, NH, NJ, NY, OH, TN, and VT. The state road networks ranged in size from Hawaii, with only 64892 vertices and 76809 edges to New York, with 716215 vertices and 897451 edges. They also varied greatly in terrain, urban areas, and presence of large areas of wilderness with sparse roads. Multiple POIs of the same type at the same address were combined into one site. However POIs in close proximity but at different vertices were treated separately.

For our two "colors" of sites, we report on experiments using *educational institutions* and *religious institutions* accessed from a publicly available collection of POIs. The number of sites in a file ranged from a minimum of 144 (educational institutions in HI) to a maximum of 7640 (educational institutions in TN). In addition to being publicly available POIs, these also made a good choice because they are intuitively distributed in a way that could lead to poor performance. Educational institutions—unlike, for example, post offices or fire stations—are unevenly distributed; a large campus for a single institution may contribute to the POI file numerous buildings in close proximity but with different addresses.

We report first on the *doubling densities* of these POIs on road networks with respect to teach other, and on then the depths to which the SSSP searches need to go before they can be pruned by Lemmas 4 and 8.

## 5.1 Doubling Density

As noted above, the preprocessing in Step 1 of the dynamic variant of the algorithm must compute a table of distances between pairs of sites that define potential Voronoi regions. In the worst case, this may take $O(km + kn \log n)$ time to compute and additional $O(k^2)$ space to store the table, where $k$ is the number of sites. However by Lemmas 3 and 7, we only need to store pairs of sites $(s, t)$ if $s$ is no more than twice as far from $t$ as the nearest other site to $t$, or vice versa. This improved efficiency for Step 1 thus depends on a property we call the *doubling density*, which is defined as follows: *for a given vertex $v$ and set of sites $S$, the doubling density of $v$ is the number of sites in $S$ no further from $v$ than twice the distance to the nearest other site to $v$ (not counting $v$ if $v \in S$.)*

We computed the average doubling density from religious institutions to educational institutions, and vice versa, as well as the number of times edges were visited on all of these searches. We report here on the key factors, which are the total number of two-color pairs of candidate sites generated by both searches—that is, the number of possibly pairs that could have non-empty Voronoi regions and are stored in the table in Step 1—and also the total number of times edges were visited in both sets of searches to compute this table

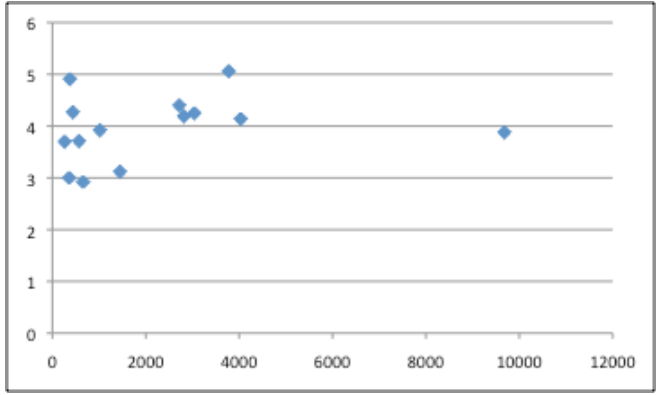

Figure 5: Values of c/k (candidate pairs divided by number of sites) as a function of k (number of sites) for fourteen states.

Let $c$ be the total double density—that is, the number of "candidate pairs" of sites that might have non-empty Voronoi regions. In the worst case, $c$ could be $\Omega(k^2)$ where $k$ is the number of sites. However empirical results for these POIs on fourteen states shows that $c$ is $O(k)$. In particular, the ratio $c/k$ of the total number of candidate pairs to the total number of sites (of both types) ranged from 2.92 to 5.06. A graph of all fourteen states is shown in Figure 5.

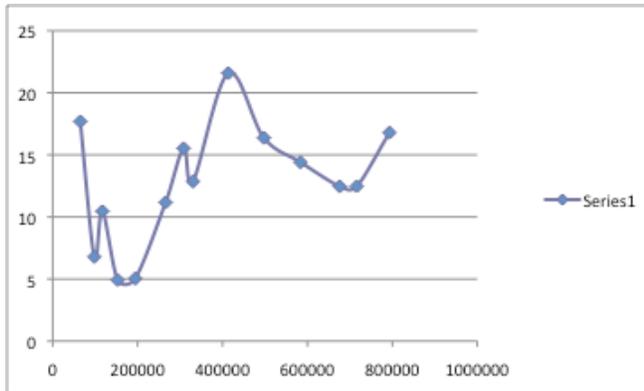

Figure 6: Based on the doubling density, the average number of times each edge is visited (in the search for candidate pairs) visited as a function of $n$ (for fourteen states.)

Furthermore, closely related to this the doubling density, the average number of times each edge is visited in computing this table, in the total of both types of searches, was less than 22 in all trials. See Figure 6. Thus empirical results suggest a constant average doubling density, a table of candidate pairs that is linear in the number of input sites, and a total run time of $O(m + n \log n)$ and total space of $O(n + m)$ for all the SSSP searches to compute this information.

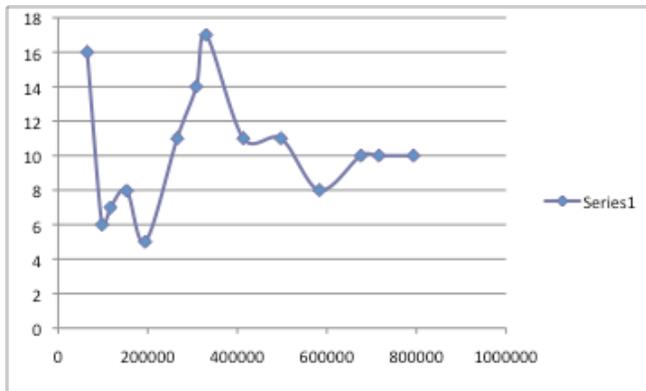

Figure 7: The average number of candidate pairs of sites *per vertex* (for fourteen states.)

## 5.2 Dynamic Pruning on Road Networks

Results on the level of pruning are equally promising, though less immediately so; an amortized approach is required to see the efficiency. In particular, when sites of one type are much denser than sites of another—as is the case, for example, when there is a large educational campus that contributes numerous entries to a POI file in a small area—then the number of pairs of sites satisfying Lemma 1 and 5 may be large, prohibiting an early pruning of the searches and requiring distance calculations for numerous pairs of sites.

However in the two-color variant, there is no second phase of the algorithm when we must perform a SSSP search from each vertex with incomplete information. Instead, the SSSP searches from the original sites continue until all of them have been pruned. So we can bound the overall run time of these searches simply by the total number of times that a search continues through a vertex—or, equivalently, by the average number of times that each vertex is visited. In all trials except Hawaii, the average number of such vertex visits was less than 10 per vertex, which is linear on the size of the graph. (For Hawaii, which had the smallest graph, the number was 16.) That is, each vertex was visited an average of $O(c)$ SSSP searches for a total of $O(n)$ vertex visits, before further searches are dynamically pruned by Lemma 8.

Equally importantly for the efficiency of the algorithm, the number of pairs of sites of different colors that are are tested for each vertex empirically appears to converge to 10 as the number $n$ of vertices grows, as shown in Figure 7. Thus, empirically the total number of distances explicitly computed to find the minimum was $O(n)$.

This immediately implies that the space complexity of all the SSSPs never exceeds $O(n)$. So empirical data suggests that Step 2 also requires $O(m+n\log n)$ time and $O(n+m)$ space.